\documentclass[prl,twocolumn,showpacs,superscriptaddress,aps]{revtex4-1}
\usepackage{graphicx,amssymb,amsmath,psfrag,color}
\begin{document}
\definecolor{darkgreen}{rgb}{0,0.5,0}
\newcommand{\be}{\begin{equation}}
\newcommand{\ee}{\end{equation}}
\newcommand{\jav}[1]{{#1}}

\title{Statistics and dynamics of the center of mass coordinate in a quantum liquid}

\author{Bal\'azs D\'ora}
\email{dora@eik.bme.hu}
\affiliation{Department of Theoretical Physics and MTA-BME Lend\"ulet Topology and Correlation Research Group, 
Budapest University of Technology and Economics, 1521 Budapest, Hungary}
\author{Bal\'azs Het\'enyi}
\affiliation{Department of Theoretical Physics and 
BME-MTA  Exotic  Quantum  Phases  Research Group,   Budapest  University  of  Technology  and  Economics,  1521  Budapest,  Hungary}
\author{C\u at\u alin Pa\c scu Moca}
\affiliation{Department of Theoretical Physics and 
BME-MTA  Exotic  Quantum  Phases  Research Group,   Budapest  University  of  Technology  and  Economics,  1521  Budapest,  Hungary}
\affiliation{Department  of  Physics,  University  of  Oradea,  410087,  Oradea,  Romania}

\date{\today}

\begin{abstract}
Motivated by recent experiments in ultracold gases, we focus on the properties of the center of mass coordinate of an interacting one dimensional Fermi gas, displaying 
several distinct phases.
While the variance of the center of mass vanishes in insulating phases such as phase separated and charge density wave phases, it remains finite in the metallic phase,
which realizes a Luttinger liquid.
By combining numerics with bosonization, 
we demonstrate that the autocorrelation function of the center of mass coordinate is universal throughout the metallic phase. It exhibits persistent oscillations and its 
short time dynamics reveal important features of the quantum liquid, such as the
Luttinger liquid parameter and the renormalized velocity.
The full counting statistics of the center of mass follows a normal distribution \jav{already for small systems}.
Our results apply to non-integrable systems as well and are within experimental reach for e.g. carbon nanotubes and cold atomic gases.
\end{abstract}


\maketitle

\paragraph{Introduction.}
Strong correlations in combination with quantum mechanics in reduced dimensions have already provided a plethora of fascinating phenomena\cite{giamarchi,nersesyan}, 
including spin-charge separation, charge fractionalization, Wigner crystals  and non-Fermi liquid behaviour.
Many of these pop up  in a variety of fermionic and bosonic systems, including condensed matter, cold atomic systems\cite{cazalillarmp}, quantum optics\cite{changnatphys} and even in black holes\cite{balasll}.
Not only compelling, but these systems promise to be relevant for possible application in topological quantum computation, spintronics and quantum information theory.

In classical mechanics, the concept of the center of mass coordinate plays a prominent role. Due to Newton's third law, the action and reaction forces between the particles
compensate each other, and the center of mass is influenced only by external forces.
The very same program can be also carried out in quantum mechanics  and the center of mass coordinate gets separated from the relative ones\cite{messiah}. 
However, this works only
when the interaction depends on the relative position of the particles and not on their absolute position. In any realistic setting in condensed matter or cold atomic systems, 
an atomic or trapping potential
is inevitably present, involving the absolute position of particles. Therefore, the center of mass contribution cannot be separated from the rest and its properties 
are influenced by strong correlations. 
Understanding how this happens is the main goal of this work, and low dimensional quantum systems featuring enhanced correlation effects represent an ideal playground for that.

The proper definition of the many-body position or center of mass coordinate has a long history\cite{resta,restasorella}, especially with periodic boundary conditions.
With open boundary conditions (OBC), however,  one can legitimately define the position operator in the conventional way\cite{rigolshastry} as $\sum_ix_i$ by summing 
over the position
operator of each particle. Moreover, experimental realizations often imply OBC. In this context, a recent experiment on weakly interacting bosons in one 
dimension has already investigated the dynamics of the center of mass\cite{geiger}.
Our aim is to shed light on the complementary, strongly correlated side of the problem, thus we focus on a strongly interacting one dimensional quantum liquid 
in one dimension with OBC\cite{cazalillaboson}.
We find that the center of mass coordinate reveals universal behaviour and its variance vanishes in insulating phases. \jav{In the Luttinger liquid (LL) phase, 
its variance gives a direct measure of the LL parameter. Its
temporal dynamics follow a universal scaling function, and reveals the other relevant parameter of the low energy theory, the renormalized velocity.}
The full counting statistics of center of mass obey a normal distribution \jav{already for small systems with 8 - 10 particles}. 
The observation of all these features is within experimental reach.

\paragraph{Interacting fermions in 1D: lattice and continuum.}
We study one dimensional spinless fermions in a tight-binding chain with nearest neighbour interaction at half filling and \emph{open} boundary
condition (OBC)\footnote{\jav{The OBC arises from a box trapping potential by using holographic masks\cite{bakr} or by confining cold-atoms in a ring lattice and cutting it open
or by using gates to confine electrons in carbon nanotubes.}} using several numerical techniques. This problem is
equivalent to the 1D Heisenberg XXZ chain after a Jordan-Wigner transformation\cite{giamarchi,nersesyan}.
The Hamiltonian is
\begin{gather}
H=\sum_{m=1}^{N-1}\left[ \frac{J}{2} \left(c^\dagger_{m+1}c_m +c^\dagger_{m}c_{m+1}\right)+J_z n_{m+1}n_m\right],
\label{xxz}
\end{gather}
where $c$'s are fermionic operators, $n_m=c^\dagger_{m}c_m$ and $J_z$ denotes the nearest neighbour repulsion, $N$ 
the number of lattice sites and the model hosts $N/2$ fermions.  This model realizes a Luttinger liquid for $|J_z|<J$ 
\jav{and the strength of the interaction is characterized by the dimensionless } LL parameter\cite{giamarchi} $K=\pi/2[\pi-\arccos(J_z/J)]$ and renormalized velocity
$v=aJ{\pi}{\sqrt{1-(J_z/J)^2}}/{2\arccos(J_z/J)}$ with $a$ the lattice constant.
For $J_z>J$, the ground state becomes  a charge density wave through a Kosterlitz-Thouless transition with broken $Z_2$ (corresponding to even/odd lattice sites) symmetry, while
for $J_z<-J$, the ground state is phase separated through a first order phase transition, i.e. all $N/2$ fermions are "bound" together. 
This model is solved using exact diagonalization (ED) with Lanczos algorithm up to $N=26$ and by the density matrix renormalization group (DMRG) up to $N=80$.

{The low energy effective field theory  of Eq. \eqref{xxz} in the LL phase is obtained using Abelian bosonization\cite{nersesyan,giamarchi,cazalillarmp}, capturing interaction effects non-perturbatively.
Using this procedure, the LL phase of this model with OBC is mapped onto\cite{fabriziogogolin,cazalillaboson}}
\begin{gather}
H=\sum_{q>0}\omega(q)b^\dagger_qb_q,
\label{hboson}
\end{gather}
where { $b_q$ accounts for the density fluctuations\cite{giamarchi} of the fermions in Eq. \eqref{xxz}} and the long wavelength part of the local charge density is $\rho(x)=\partial_x\Theta(x)/\pi$
with
\begin{gather}
\Theta(x)=i\sum_{q>0}\sqrt{\frac{\pi K}{qL}}\sin(qx)\left[b_q-b^\dagger_q\right]
\label{thetax}
\end{gather}
for OBC
and $K$ the LL parameter and $\omega(q)=vq$ with $v$ the Fermi velocity in the interacting systems and $q=l\pi/L$ with $l=1,2,3\dots$.

\paragraph{Center of mass.}
We define the dimensionless center of mass operator for Eq. \eqref{xxz} as\cite{vaugham}
\begin{gather}
\hat x=\frac{1}{N}\sum_{m=1}^{N}\left(m-\sum_{m'=1}^N\frac{m'}{N}\right) n_m,
\label{xop}
\end{gather}
where for simplicity, we have subtracted the equilibrium position of the center of mass coordinate such that $\langle \hat x\rangle=0$, irrespective of how the lattice sites are numbered.
For identical particles, what we consider here, it is independent from their mass.
This operator is also  the normalized polarization operator\cite{mahan}.
Using bosonization, the very same quantity reads as
\begin{gather}
\hat x=\int_0^L\frac{dx}{\pi L} x \partial_x\Theta(x),
\end{gather}
and we have neglected fast oscillating terms in the 
integrand\cite{giamarchi,nersesyan,cazalillarmp} from short wavelength density fluctuations, which are expected to average out after the integral.

\begin{figure}[h!]
\psfrag{x}[t][][1][0]{$J_z/J$}
\psfrag{y}[b][t][1][0]{{\color{red}$10~\sigma_x^2$}, {\color{blue}$v=\dfrac{7\zeta(3)Lc_x}{\sigma_x^2\pi^3}$}}
\includegraphics[width=7cm]{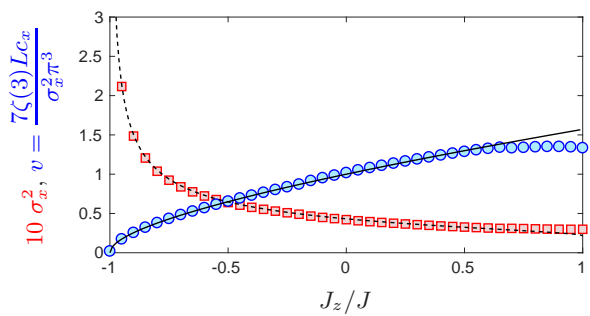}
\caption{The variance (red squares) of the center of mass and the LL velocity (blue circles) from the center of mass autocorrelator
are plotted from DMRG with $N=80$. The analytical results from Eq. \eqref{x^2} and \eqref{cx} using the Bethe Ansatz results for $K$ and $v$ {without} any fitting parameter are shown by black lines.}
\label{variancell}
\end{figure}

While the expectation value of the center of mass operator is zero, its standard deviation, $\sigma_x$ reads as
\begin{gather}
\sigma_x^2=\langle \hat x^2\rangle=\int_0^Ldx\int_0^Ldy \frac{xy\langle \partial_x\Theta(x)\partial_y\Theta(y)\rangle}{\pi^2 L^2}=\nonumber\\
=\sum_{q>0}\frac{Kq}{\pi L}\left[\int_0^Ldx\cos(qx)\frac{x}{L}\right]^2=\frac{7\zeta(3)}{2\pi^4} K,
\label{x^2}
\end{gather}
where $\zeta(z)$ is the Riemann zeta function\cite{gradstein} and $\zeta(3)\approx 1.202$.
The $\sigma_x$ is \jav{universal in the sense that it depends only on the LL parameter, $K$, but is independent of the high energy degrees of freedom:
very different microscopic Hamiltonians with the same LL parameter possess identical $\sigma_x$.}
Since $K$ decreases with increasing $J_z$, this implies that counterintuitively, the variance gets suppressed when moving from the attractive to the repulsive side.
The numerical results from DMRG agree very nicely with Eq. \eqref{x^2}, as seen in Fig. \ref{variancell}. 
The variance diverges as $\sigma_x^2\sim \sqrt{J/(J+J_z)}$ at the first order critical point. Slight deviations are visible close to $J_z\sim J$, arising from 
the terms in the Hamiltonian, driving the Kosterlitz-Thouless transition, which are missing from Eq. \eqref{hboson}.
Nevertheless, the variance seems to remain finite at this critical point.

The above calculation can be extended to the gapped charge density wave phase, when the effective field theory of Eq. \eqref{xxz} is the sine-Gordon model\cite{giamarchi,nersesyan}.
In this case, a Mott gap $\Delta$ opens up in the spectrum. Within the realm of the semiclassical limit of this model, following Ref. \onlinecite{maki79,iuccisinegordon},
the variance of the center of mass is calculated with $K$ replaced by $\omega(q)/\sqrt{\omega^2(q)+\Delta^2}$ in Eq. \eqref{x^2} under the sum.
This gives  $\sigma_x^2\sim v/L\Delta$ and vanishes in the thermodynamic limit, which  is also corroborated by ED.
For finite systems, the variance vanishes when the system size, $L$ is much longer than the correlation length, $v/\Delta$.  Alternatively,
the variance is negligible when the level spacing, $v/L$ is much smaller than the actual gap\footnote{This would allow for estimating the gap size from measuring
the variance of the center of mass in finite size systems.}.

In the phase separated regime, bosonization is not applicable, but the variance of the center of mass can be calculated. Since all $N/2$ particles are bound together by
the strong attractive interaction in the lattice of $N$ sites, the ground state is in principle highly degenerate. As a result, $\sigma_x\sim N/2$, 
which agrees with ED results on clean systems.
However, any disorder or imperfection in the lattice, which is inevitably present in any real system, breaks this degeneracy and produces a unique ground state.
Therein, the $N/2$ particles occupy neighbouring lattice sites, their position is well defined and the variance is zero, as we also find from ED in the presence of weak impurities or disorder.

\begin{figure}[h!]
\psfrag{x}[t][][1][0]{$t/t^*$}
\psfrag{y}[b][t][1][0]{$\chi_x(t)/\sigma_x^2$}
\includegraphics[width=7cm]{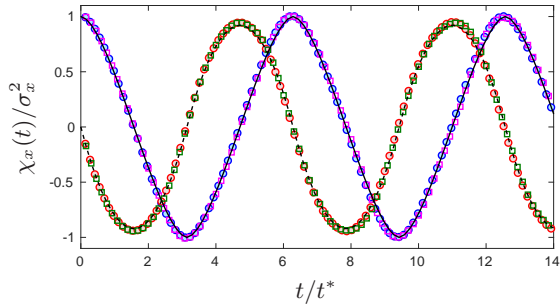}
\caption{Demonstration of the data collapse for LL with strong attraction and repulsion. The real (black solid) and imaginary (black dash-dotted ) part of the center of
mass autocorrelator is plotted from Eq. \eqref{xx(t)} with $t^*=L/v\pi$, together with the numerical data from ED
for $N=26$ and $J_z/J=0.6$ (circle) and -0.6 (square), normalized the to corresponding variance.  The only fit parameter for ED
is the horizontal timescale, satisfying $t^*\sim L/v\pi$.
\label{xxt}}
\end{figure}

\paragraph{Dynamics of the center of mass.}
To gain further insight into the behaviour of the center of mass operator, we evaluate its autocorrelation function as
$\chi_x(t)=\langle \hat x(t)\hat x(0)\rangle$.
Using $b_q(t)=b_q\exp(-i\omega(q)t)$ in Eq. \eqref{thetax}, we
obtain
\begin{gather}
\chi_x(t)=\frac{2K}{\pi^4} \sum_{l=1}^\infty \frac{1-(-1)^l}{l^3}\exp(-ivt\pi l/L)=\nonumber\\
=\frac{2K}{\pi^4}\sum_{b=\pm}b~\textmd{Li}_3(b~\exp(-ivt\pi/L))
\label{xx(t)}
\end{gather}
with Li$_s(z)$ the polylogarithm function, and gives $\chi_x(0)=\sigma_x^2$. Although Eq. \eqref{xx(t)} looks complicated at first, it is rather well approximated by
$\chi_x(t)\approx\sigma_x^2\exp(-ivt\pi/L)$.
Similarly to the variance of the center of mass, $\chi_x(t)$ is also independent of any cutoff and depends only on the universal combination $vt/L$.
Its initial temporal slope is
\begin{gather}
c_x=i\partial_t\chi_x(t\rightarrow 0)=\langle [\hat x,H]\hat x\rangle= \frac{1}{2\pi L} vK,
\label{cx}
\end{gather}
which depends only the the LL parameter  $K$ and the renormalized velocity of the interacting theory.
Therefore, by measuring the variance of the center of mass and its initial dynamics, one can easily extract the two and only two essential ingredients 
of the LL theory, the velocity from $v=7\zeta(3)Lc_x/\sigma_x^2\pi^3$ and the LL parameter from $\sigma_x$, as shown in Fig. \ref{variancell}.
In addition, Eq. \eqref{xx(t)} predicts a universal data collapse of the center of mass oscillation, namely upon rescaling its magnitude by $1/K$ and
its temporal evolution by $v/L$, all curves should fall on top of each other, irrespective of the strength or even the sign of the interaction, as shown in Fig. \ref{xxt}.
The time dependence spans several $N/J$ periods (with $N=26$) and the agreement between numerics and Eq. \eqref{xx(t)} remains excellent, even
though $K$ and $v$ decreases/increases by more
than a factor of 2 from $J_z/J=-0.6$ to 0.6, respectively.

The center of mass autocorrelator is found to be universal at all timescales. This is somewhat surprising since the LL theory is designed to capture the low energy physics,
thus it is expected to be universal in the long time limit. For $\chi_x(t)$, on the other hand, already the short time dynamics turns out to be universal.
The lattice model in Eq. \eqref{xxz} in integrable\cite{giamarchi,nersesyan} therefore one may wonder whether these persistent oscillation arise due to the large number of constants of motion.
Integrability is destroyed by adding a second nearest-neighbour  density-density (i.e. $J_z' \sum_{m} n_{m+2}n_m$) interaction\cite{hallberg}, what  we have also studied numerically for several $J_z$ and $J_z'$, yielding identical results to the integrable case:
the persistent oscillations from Eq. \eqref{xx(t)} remain intact also for non-integrable LLs.

Persistent oscillation shows up in the Calogero-Sutherland model\cite{sutherland} as well, sensitive to the trapping frequency.
This is argued to be a specific feature following from the integrability of the model and its long range interaction.
The persistent oscillation in Eq. \eqref{xx(t)} is analogous to this and scales with the "trapping frequency" $v\pi/L$ from OBC.
\jav{The OBC can also arise from a sharp box trapping potential\cite{bakr,gaunt}.}
However, the LL description applies to a large variety of systems, including fermions, bosons, spins\cite{cazalillarmp} etc. 
Therefore the persistent oscillation is expected to be a generic feature in these models, irrespective
of the microscopic details.

\paragraph{Full counting statistics.}
Already
simple expectation values of physical quantities
often display rather complex behavior.
Higher moments of the observables contain, however, infinitely more information
and encode unique information about e.g. 
non-local, multi-point correlators  and entanglement, though
they are typically difficult to access.
Their information content is equivalent to determining the full distribution function of the quantity of interest.

Having studied simple correlation functions of the center of mass coordinate, we now address its full counting statistics\cite{lesovik,gritsev,silva,gring}.
Its probability distribution function is
\begin{gather}
P(X)=\langle\delta(X-\hat x)\rangle,
\end{gather}
whose characteristic function can easily be evaluated to yield
$G(p)=\langle\exp(i\hat xp)\rangle$.
Note that $G(p)$ is reminiscent to how the polarization operator is defined\cite{resta,restasorella,kobayashi} for periodic boundary condition,
using only integer multiples of $2\pi$ for $p$. Here, on the contrary, $p$ takes any real values in the characteristic function and
 the normalized position and polarization operator, $\hat x$ is defined by Eq. \eqref{xop} without any ambiguity\cite{rigolshastry}.

\begin{figure}[h!]
\psfrag{x}[t][][1][0]{$p\sigma_x$}
\psfrag{y}[b][t][1][0]{$G(p)$}
\includegraphics[width=7cm]{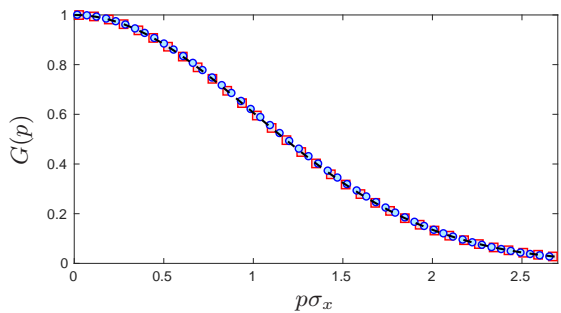}
\caption{Characteristic function of the center of mass coordinate from bosonization (dashed line) together with the numerical data from ED after finite size scaling to $N\rightarrow\infty$
using $N=14$, 18, 22 and 26  and $J_z/J=0.6$ (circle) and -0.6 (square).  
\label{xpdfll}}
\end{figure}

Since $\hat x$ in the exponent is a linear function of bosonic operators, and the low energy Hamiltonian is quadratic in Eq. \eqref{hboson},
the expectation value is evaluated as\cite{delft} 
\begin{gather}
G(p)= \exp(-p^2\langle \hat x^2\rangle/2)=\exp(-p^2\sigma_x^2/2).
\label{charac}
\end{gather}
This is calculated  also for Eq. \eqref{xxz} numerically using ED after finite size scaling, and plotted in Fig. \ref{xpdfll}, revealing excellent agreement between 
Eq. \eqref{charac} and the numerical data. For smaller systems and especially for repulsive $J_z\sim J$, slight deviations show up from the Gaussian behaviour for large $p$, which
stem from the fact that $\| \hat x\|< N/2$ is bounded for finite systems, therefore deviations appear in the tail, which diminish upon increasing the system size.
Its Fourier transform gives the probability distribution function as a normal distribution with variance $\sigma_x^2$ as
$P(X)=\frac{1}{\sqrt{2\pi}\sigma_x}\exp\left(-\frac{X^2}{2\sigma_x^2}\right)$.
The normal distribution itself is expected from the central limit theorem in the thermodynamic limit.
However, on the one hand, 
it is surprising that already for small system sizes, where higher moments could in principle deviate from gaussianity, the numerical data approaches it very fast 
already for systems with 8 - 10 particles, especially for attractive interactions.
The same distribution applies  for attractive Bose-Einstein condensates in a harmonic trap\cite{sakmann}.
On the other hand, in contrast to the smooth, almost $N$ independent behaviour of $G(p)$ for OBC and its nice agreement with bosonization in Fig. \ref{xpdfll},
the very same quantity exhibits power law size dependence for \emph{periodic} boundary condition 
as $G(p)\sim N^{-\alpha(p)}$ and the exponent $\alpha(p)$ does not follow the field theory prediction\cite{kobayashi}.

\paragraph{Experimental ramifications.}
There exists  well-developed  experimental technology to observe these effects.
\jav{LLs are routinely realized in both cold atomic settings, using spins, bosons or fermions,  and condensed matter systems\cite{cazalillarmp}, including e.g. carbon nanotubes,
described identically by Eq. \eqref{hboson}}.
The center of mass
coordinate can be monitored using time of flight imaging\cite{BlochDalibardZwerger_RMP08}, in-situ absorption imaging\cite{geiger} 
 or scanning tunneling microscopy\cite{ilani}, allowing for the observation
of its variance as well as its full distribution function, or at least some of its lower moments. These are all universal quantities, depending on the interaction 
only through the LL parameter $K$. This is tunable by changing the lattice parameters or tuning the Feshbach resonance for cold atoms in a wide range, while
the interaction in condensed matter is controllable by tuning the relative permittivity of the surrounding material.

The dynamics of the  center of mass coordinate is measurable by e.g. tilting the lattice or applying a weak electric field at time $t=0$, 
represented by the scalar potential of the force $F$, which creates a perturbation $H'=L\hat xF(t)$ as in Ref. \cite{dahan,gustavsson}.
Then, within linear response theory, the motion of the center of mass follows as
\jav{\begin{gather}
\langle \hat x(t)\rangle=-2L\int_0^t dt'\textmd{Im}\chi_x(t-t')F(t').
\label{xtf}
\end{gather}}
For short times, $\textmd{Im}\chi_x(t)= -vKt/2\pi L$, revealing the two LL
characteristics in a universal manner. Therefore initially $\langle \hat x(t)\rangle=vKt^2F/2\pi$ after switching on a constant force $F$, 
corresponding to the classical motion of a particle in an external force $F$ with "mass" $\sim \pi/vK$.
Based on Eq. \eqref{xx(t)} and Fig. \ref{xxt}, the $\langle \hat x(t)\rangle$ will exhibit persistent oscillations for longer times with frequency $v\pi/L$. 
At the same time, the variance of the oscillating center of mass remains unchanged and it does not spread during the oscillations, in spite of being built up from many distinct 
dispersive modes. 
\jav{Note that Eq. \eqref{xtf} is \emph{exact} within the realm of bosonization, there are no higher order corrections in $F$. This
follows from the linear dispersion in $\omega(q)\sim q$,
extending up to infinitely large energies, without any band bending. This is completely analogous to
how the Born scattering limit of Dirac-delta potential
is exact for the same linear dispersion\cite{nersesyan}.}

The experimental setup  in Ref. \onlinecite{geiger} can be readily used to investigate these predictions. Therein, a weakly interacting Bose-gas of $^7$Li was monitored and the dynamics of its center of mass was measured
in the presence of strong driving force, while our results apply in the opposite case of strong interaction and weak driving field, which is realizable experimentally. 
In a related experiment\cite{fujiwara}, the center of mass of non-interacting $^7$Li particles was measured in an excited band, but interactions can be induced by making use of its Feshbach resonances\cite{chin}.
Our fermionic model in Eq. \eqref{xxz} can equally be realized  in terms of hard core bosons\cite{giamarchi}, which corresponds to the Tonks-Girardeau limit of a 1D Bose gas~\cite{paredes}, created from  $^{87}$Rb. The dynamics of the 
center of mass is accessible following Refs. \cite{geiger,fujiwara}.

\jav{The dynamics of the center of mass
is reminiscent to Bloch
oscillations\cite{davies,dahan}, which also arise in the presence of an external force, }
albeit the persistent oscillations in Eqs. \eqref{xx(t)} and \eqref{xtf} arise in a strongly correlated quantum liquid as opposed to the standard single particle picture
behind Bloch
oscillations\cite{davies}.
The analogy with Bloch oscillation is extended  by noting that the reflection on the boundary of the lattice in our study plays the role of 
Bragg reflection \jav{at the boundary of the Brillouin zone in the case of Bloch oscillations}.
The typical 
timescale of Bloch oscillations, $t_B=1/aF$ with $a$ the lattice constant, represents the time during which the full Brillouin zone is swept through by the force, while 
the timescale for the center of mass oscillation due to finite size effects from Eq. \eqref{xx(t)} is
$L/v$, i.e. the timescale for sweeping through the real space lattice. \jav{Our results are observable on the timescale of $t\sim L/v\ll t_B$ before Bloch oscillations set in,
requiring weak forces in Eq. \eqref{xtf} and more importantly,
small systems, which suits ideally the experimental conditions.}

\paragraph{Conclusions.}
\jav{We have demonstrated that the  center of mass coordinate exhibits  universal behaviour  in a Luttinger liquid
and bosonization gives essentially exact resuls for \emph{all} of its properties.
Most importantly, the LL parameter can be directly measured using the variance of the center of mass coordinate. 
In combination with its short time dynamics, the other basic characteristic of the underlying quantum liquid, namely the renormalized velocity, is revealed.} 
\jav{The correlation function as well as the full counting statistics of the center of mass coordinate}
follow a universal function, which are corroborated by analytical and numerical methods.
These are within experimental reach both in condensed matter and cold atomic realizations, using setups similar to Bloch oscillations.

\begin{acknowledgments}
We are grateful for useful discussions with I. Lovas.
This research is supported by the National Research, Development and Innovation Office - NKFIH   within the Quantum Technology National Excellence Program (Project No.
      2017-1.2.1-NKP-2017-00001), K119442 and by
 UEFISCDI, project number PN-II-RU-TE-2014-4-0432.
\end{acknowledgments}

\bibliographystyle{apsrev}
\bibliography{wboson1}

\end{document}